\documentstyle[12pt,aps,preprint,epsf]{revtex}

\preprint{UCRHEP-T290}

\def\pperp{{\mbox{\raisebox{-.05cm}{$\stackrel{\perp}{p}$}}}{}}
\def\up#1{^{(#1)}}
\def\lsim{\,{\raise-3pt\hbox{$\sim$}}\!\!\!\!\!{\raise2pt\hbox{$<$}}\,}
\def\gsim{\,{\raise-3pt\hbox{$\sim$}}\!\!\!\!\!{\raise2pt\hbox{$>$}}\,}

\begin{document}

\title{Mass dependence of the gravitationally-induced \\ wave-function phase}
\author{ Jos\'{e} Wudka }
\address{Department of Physics \\
University of California-Riverside \\
California, 92521-0413\\
}
\date{\today}
\maketitle

\begin{abstract}
The leading mass dependence of the wave function phase is
calculated in the presence of gravitational interactions. The conditions
under which this phase contains terms depending on both the square of
the mass and the gravitational constant are determined. The
observability of such terms is briefly discussed.
\end{abstract}

\bigskip

\bigskip\bigskip

\paragraph{Introduction}
There has been a series of publications discussing the mass dependence of
the gravitational effects on the phase of the wave
function~\cite{al,mot,oth}, and, due to
a divergence in the results, a certain amount of controversy has
resulted. Some results~\cite{al} indicate the presence of a
gravitationally-induced phase proportional to the square of the masses,
while other calculations show that such terms are in fact absent~\cite{mot}, 
and that the leading
dependence in the gravitational contributions to the phase is proportional
to the fourth power of the mass. In this short note I will attempt to
resolve this (apparent) contradiction. I will first consider the case of 
neutrinos propagating in an arbitrary gravitational field and then
generalize to other particles

\paragraph{Neutrinos in a gravitational field}
Using a WKB approximation it is possible to derive the effective
Hamiltonian for neutrinos propagating in a non-trivial metric. Following
the procedure described in detail in~\cite{prw} the relevant expression
is~\footnote{The Hamiltonian has units of mass$^2$ due to the choice of
evolution parameter, see below}
\begin{equation}
H_{\rm eff} = {1\over4} p_a \left( \partial_\nu e_{b\mu} \right) e^\mu_c e^\nu_d
\left[ \epsilon^{a b c d} - 2 \eta^{ a b} \epsilon^{c d e f} { p_e
\pperp_f \over \pperp^2} \right] \gamma_5 + {1\over2} m^2,
\label{heff}
\end{equation}
where Latin indices refer to the local Lorentz frame, Greek indices to
the global coordinate frame, and $e^\mu_a$ denote the tetrads. 
$p$ is the momentum of a null geodesic which the neutrino wave packets 
follow, within the WKB approximation, when $ m=0 $. I will denote by $\lambda$
the affine parameter along this geodesic
\begin{equation}
p^\mu = { d x^\mu \over d\lambda } .
\end{equation}
The vector $ \pperp$
denotes the time component of $p$ with respect to a comoving reference
frame.~\footnote{More precisely: denoting by $ \nu^\mu_A$, $A=1,2,3$ three
independent (local) solutions to the geodesic deviation
equation~\cite{weinb}, $\pperp$ is the component of $p$ orthogonal 
to the $ \nu^\mu_A$.} The  Hamiltonian $ H_{\rm eff} $ generates 
translations in $\lambda$.

Though the pure gravitational terms in (\ref{heff})
can generate interesting effects, here I
will concentrate on the contributions generated by the term $ m^2/2 $.
It is clear from the above expression that each mass eigenstate acquires
a phase equal to 
\begin{equation}
\Phi\up m = - {1\over2} m^2 \lambda.
\label{phase}
\end{equation}
There are subleading contributions to $H_{\rm eff} $ which depend on the
mass~\cite{prw}, but these are of order $ m/R $ where $R$ denotes the distance
scale of the metric, and can be ignored for $ m > 10^{-11}$eV.

\paragraph{Other particles}
The same expression for the mass dependence of the phase can be obtained
using the following argument. Within the WKB approximation the wave
function of a general quantum system takes the form 
\begin{equation}
\Psi = e^{i S} \chi ,
\end{equation}
where $S$ denotes the classical action and $ \chi$ a slowly-varying
amplitude. When the particle moves in a
non-trivial metric background $S$ satisfies the following
Hamilton-Jacobi equation~\cite{ll}
\begin{equation}
g^{\mu\nu} \partial_\mu S \partial_\nu S = m^2,
\label{hj}
\end{equation}
(I ignored all other interactions and considered a single
mass eigenstate). The action is generally covariant so that the
corresponding phase is unambiguously determined (within the semiclassical
approximation)

For small $m$ the action takes the form
\begin{equation}
S = S_0 + m^2 S_1 + \cdots, 
\label{action}
\end{equation}
which, when substituted into (\ref{hj}) yields
\begin{equation}
g^{\mu\nu} p_\mu p_\nu = 0 , \qquad  p^\mu \partial_\mu S_1 = - {1\over2},
\end{equation}
where
\begin{equation}
p^\mu = - g^{\mu\nu} \partial_\nu S_0 = { d x^\mu \over d\lambda},
\label{tangent}
\end{equation}
is the tangent vector to the null geodesic. It then follows that 
$ p^\mu \partial_\mu S_1 =  d S_1/ d \lambda $,
so that $ S_1 = - \lambda m^2 /2 $ which gives the same contribution as
(\ref{phase}). It follows that this result is general provided $m$ is
small compared to the particle's momentum.

\paragraph{Observability of the gravitationally induced phase}
\label{sect:observability}
I will consider a wave packet that can be decomposed into a coherent
superposition of mass eigenstates. Using standard arguments the above 
expressions imply that, within the semiclassical approximation, the phase 
$ S_0 + \lambda m^2/2 $ may lead to  oscillations between components of
different mass. This will occur provided {\sl (a)}
the amplitude of the mass components of the state have a sufficient overlap at the
observation point and, {\sl (b)} the phase difference between them
is sufficiently large ($ \gsim \pi $). I will first
review briefly the conditions for under which there is significant
amplitude overlap and then discuss the effects of the phase difference.

If the mass components are created in the {\em same} localized space-time
region $A$ then, for long distances and within the WKB approximation,
they will overlap at the observation region $B$ only if the
corresponding amplitudes in momentum space are centered at different
momenta. In this
case the phase difference is simply $ \Delta m^2 ( \lambda_B - \lambda_A
)/2 $.

Alternatively we can assume that the initial mass amplitudes are
localized in the same region of momentum space. Then, in order for them
to arrive simultaneously at $B$ (so that there is significant amplitude overlap), 
they must originate at different space-time
locations $A$ and $A'$, which may be space or tiime-like separated 
(the creation region then contains
$A$ and $A'$ and is sufficiently large to allow a small momentum
uncertainty). In this case the
phase difference is $ S_0(A)-S_0(A')+\Delta m^2 ( \lambda_B - \lambda_A
)/2 $, where $ S_0(A)-S_0(A')$ is also of order $ \Delta m^2 $.
This is the scenario described in~\cite{mot}.

Leaving aside the issue of which of the two scenarios above is more
natural, it is clear that the expressions for the phase difference will be
different in each case because the assumed initial conditions are different~\cite{oth}.

Even if interference effects are present, an
{\em independent} question is whether the phase difference between two mass
eigenstates has a strong dependence on the gravitational field. I will consider 
for simplicity the first case described above,
where the phase for each mass component is given by (\ref{phase}) and depends only
on the geodesic length $\lambda $. In order to determine
the dependence on the gravitational force this expression should be re-written in terms
of a set of physical observables, but such a set is no unique and different choices
may exhibit varying dependences on the gravitational constant.

The expression for $\Phi\up m$ will
depend on the geodesic parameters such as the energy and angular momentum. 
If, for example, the interference effects are measured by a freely-falling
observer, these should be written in terms of the corresponding quantities
measured by the observer (e.g. the energy-momentum vector equals $e^a_\mu p^\mu$).
In addition, this phase will depend on the physical distance form the source to the
observer. This distance can be
defined as half the time it takes for light to go from $A$ to $B$ and
back to $A$ (in units where the speed of light is one); explicitly~\cite{ll}
\begin{equation}
\ell_{AB} 
= \int_A^B d\lambda \sqrt{ \left( g_{ij} - g_{0i} g_{0j}/g_{00} \right) p^i p^j }   
= E \int_A^B {d \lambda \over \sqrt{-g_{00}} }
\label{inv.l}
\end{equation}
(Latin indices denote the spatial components) the second expression 
is valid for null geodesics in a static
space time (the integrals are evaluated along the geodesic). 
Note however, that $\ell_{AB}$ is {\em not} the only possible quantity
that can be associated with the physical distance from the source to the
observer. 
Other possibilities are in general available and, as will be 
illustrated below, the form of
$\Phi\up m $  are strongly dependent on the choice made.

There is one final ``experimental'' issue concerning the observability 
of the phase difference. In many
situations the distance to the source is only approximately known
and the location of creation region is known only with limited accuracy. In such cases
the gravitational effects will be distinguishable only if the phase
difference due to this uncertainty can be neglected or otherwise taken into account.

\paragraph{Gravitational dependence of the quantum phase in a Kerr metric.}
The expression (\ref{phase}) contains a dependence on the gravitational
interactions due to the non-trivial expression of $ \lambda $ in terms
of the coordinates. As an illustrative example I will consider the case
of  Kerr metric, 
\begin{equation}
ds^2 = \left(1 - { r r_g \over \rho^2} \right) dt^2 - { \rho^2 \over
\Delta} dr^2 - \rho^2 d\theta^2 - \left( r^2 + a^2 + { r r_g a^2 \over
\rho^2} \sin^2\theta \right) \sin^2 \theta d\phi^2 + { 2 r r_g a \over
\rho^2} \sin^2\theta d\phi dt,
\label{kerr}
\end{equation}
where
\begin{equation}
\Delta = r^2 - r r_g + a^2 , \qquad \rho^2 = r^2 + a^2 \cos^2\theta.
\label{rd}
\end{equation}
The gravitational radius $r_g $ equals $ 2 M G $ where $M$ is the mass of the
black hole and $G $ Newton's constant. The black hole angular momentum
equals $ a M $.

In this case the Hamilton-Jacobi equation (\ref{hj}) with $m=0 $ is
separable,
\begin{equation}
S_0 = - E t + W_r(r) + W_\theta(\theta) + L \phi,
\end{equation}
where $E$ denotes the energy and $L$ the (azimuthal) angular momentum for
the particle (the $z$ axis is along the the black hole rotation axis).
Both $E$ and $L$ are constant, the addition constant of
the motion $K$ is related to the total angular momentum. Substituting,~\cite{ll}
\begin{eqnarray}
\left( { d W_\theta \over d \theta }\right)^2  = K - \left( a E
\sin\theta - { L \over \sin \theta} \right)^2 , \qquad
\left( { d W_r \over d r }\right)^2  = { [ (r^2 + a^2 ) E - a L ]^2 -
K \over \Delta^2 } .
\end{eqnarray}
Using (\ref{tangent}) it follows that
\begin{equation}
\lambda = \int^r { \rho^2 \; dr \over \sqrt{[ (r^2 + a^2 ) E - a L ]^2 -
K \Delta] } }
\label{ap}
\end{equation}
with $ \rho $ and $ \Delta $ defined in (\ref{rd}) where $\theta$
(buried in $ \Delta$) is assumed to be expressed in terms of $r$ using
\begin{equation}
\left( { d\theta\over dr} \right)^2 = { K- ( a E \sin\theta - L
/\sin\theta)^2 \over  [ (r^2 + a^2 ) E - a L ]^2 - K  \Delta },
\end{equation}
which also follows from (\ref{tangent}). The
flat-space limit of (\ref{ap})
corresponds to $ r_g = a =0 $ and $ K = L^2 $, in this case
\begin{equation}
\lambda_{\rm flat~space} = {1\over E} \sqrt{r^2 -(L/E)^2},
\end{equation}
which is a standard result. Note that $ \lambda $ has units of
mass$^{-2}$, so that $ H_{\rm eff} $ has units of mass$^2$
as noted earlier.

To lowest order in $ r_g, ~ a, ~K - L^2 $ I find,
\begin{equation}
\lambda = {1\over E} \sqrt{r^2 -(L/E)^2}
- { (K -L^2) + 2 E L a - E^2 r r_g \over 2 E^3 \sqrt{ r^2 - (L/E)^2 } } + \cdots .
\end{equation}
In the limit of zero angular 
momentum~\footnote{ $a$ is the black-hole angular momentum per unit mass} 
$ K = L^2 =0 $, $a=0 $
this expression retains a term proportional to $r_g$, but this term is actually an
irrelevant constant, $ r_g/(2E)$. It follows that for weak gravitational fields 
and small masses (\ref{phase}) becomes
\begin{equation}
\Phi\up m = 
-{m^2\over 2E} \sqrt{r^2 -(L/E)^2}
+ { m^2 [(K -L^2) + 2 E L a ]\over 4 E^3 \sqrt{ r^2 - (L/E)^2 } }
- {m^2 r_g (r -\sqrt{ r^2 - (L/E)^2 }) \over 4 E \sqrt{ r^2 - (L/E)^2 } } 
+ \cdots ,
\label{phi.r}
\end{equation}
where I explicitly subtracted the above-mentioned constant.
It follows that $ \Phi\up m $, {\em when written in terms of $r$},
does receive
gravitational contributions, but these are proportional to an
angular momentum (squared).
This dependence on the angular momentum  significantly
suppresses the magnitude of these gravitational terms. In a region very
distant from the black hole, $ r \gg (L/E) $ the above expression
reduces to (see also \cite{oth})
\begin{equation}
\Phi\up m = 
-{m^2 r\over 2E} \left\{ 1 
+ { (K -L^2) + 2 E L a \over 2 E^2 r^2 }
+ \cdots \right\}, \qquad ( r \gg L/E).
\label{kerr.a}
\end{equation}

For zero orbital angular momentum there is still a contribution to
(\ref{phase})
proportional to $r_g$, but it is also proportional to the square of
the black-hole angular
momentum. For example, for a geodesic in the $ \theta = \pi/2 $ plane 
\begin{equation}
\Phi\up m = -{m^2 r\over 2E} \left( 1 + { a^2 \over r^2 } + \cdots
\right); \qquad (L=0~\theta=\pi/2).
\label{kerr.b}
\end{equation}

The expressions (\ref{kerr.a},\ref{kerr.b}) contain no explicit gravitational contribution
when the geodesics and the black hole have zero angular momentum.
{\em However} when expressed in terms of the invariant length (\ref{inv.l})
the phase (\ref{phi.r}) becomes
\begin{equation}
\Phi\up m= - {m^2\over2E} \left[ \ell - {r_g\over 2} {\rm arcsinh}(E\ell/L)
+ \cdots\right]
\label{kerr.c}
\end{equation}
which {\em does} contain a term $ \sim r_g m^2 \ln \ell $~\cite{al}
which persists even in the $ L \to 0 $ limit 
(note that in this limit there is a term proportional to $ \ln L $, but it
is constant and will cancel when considering phase differences).
Yet, despite appearances, (\ref{phi.r}) and (\ref{kerr.c}) are equivalent.

The calculation presented in~\cite{mot} 
was done for radial geodesics in a Schwarzschild metric and corresponds
to the result (\ref{phi.r}) when $ K = L = a =0 $, in which case there is no
explicit gravitational terms present in the phase. The {\em same}
expression corresponds to the results in~\cite{al} provided the quantity 
$ \int d{\rm L}/r $ used in these references (where L represents a distance
and should not be confused with the angular momentum as used
in this paper)  is identified with $ 2 \ln \ell
$. It is worth emphasizing that both $ \ell $ and $r$ have reasonable physical 
interpretations as
measures of the distance to the source~\footnote{For example, in the case
$a=0 $, $r$ can be defined in terms of the circumference of circles around the 
origin while the interpretation of $\ell$ was provided in paragraph
{\it\ref{sect:observability}}.} and both reduce t0 the usual flat-space
distance in the limit of zero gravitational interactions. 

In order to extract the gravitational term from these expressions one
must subtract the non-gravitational contribution. This, however,
is an ambiguous quantity. For example, an observer defining the flat space
contribution as $ \Phi\up m _{\rm flat} = - m^2 r/(2 E) $ will find a small
gravitational contribution of order $(\hbox{ang. momentum})^2/(r E)^2 $
to (\ref{phase}). A second observer might alternatively define 
$ \Phi\up m _{\rm flat} = - m^2 \ell /(2 E) $ and find a much more
significant gravitational contribution $ \sim r_g \ln \ell $
to (\ref{phase}). Equivalently, an experiment done in flat space over a
distance $ d$ will generate a phase $ - m^2 d/(2E) $, in comparing this
to a similar measurement in curved space, one must determine what
quantity ($r,~\ell,$ etc.) is to be used in lieu of $d$, and the
expression for $ \Phi\up m_{\rm flat} $
depends significantly on the choice made. In this  sense the results presented 
in~\cite{al} and~\cite{mot} are, in fact, consistent.

As mentioned in paragraph {\it\ref{sect:observability}} the
the observability of these effects is a separate issue, which I discuss
briefly in paragraph {\it\ref{sect:discussion}} below. Here I merely note 
that the phase (in terms of $r$) is of order
$(m^2/E)(r+ r_g^2/ r ) $ when $ L \sim r_g E $; the second term is
absent for the case of zero angular momentum. In terms of $\ell $ 
and the phase contains a term $ \sim ( m^2 r_g/E) \ln \ell $,
even when the angular momenta 
vanish. It is possible, however, that these simple results are consequences 
of the high degree of symmetry of the metric (\ref{kerr}). 
This possibility is investigated below.

\paragraph{General spherically symmetric metric}
The metric for this configuration is~\cite{ll,weinb}, using polar coordinates,
\begin{equation}
ds^2 = e^\nu dt^2 - e^\gamma dr^2 - r^2 ( d\theta^2 + \sin^2\theta
d\phi^2),
\end{equation}
where $ \nu$ and $ \gamma $ depend on $r$ and $t$ only. The geodesics
of this metric lie on a plane which I take as the $ \theta = \pi/2 $;
then (\ref{hj}) with $m=0 $ is again separable and its solution takes
the form $ S_0 = - E t + L \phi + W_r(r) $ where $ W_r'{}^2 = e^\gamma
(E^2 e^{-\nu} - L^2/r^2) $ so that
\begin{equation}
{d r \over d \lambda } = e^{-(\nu+\gamma)/2 } \sqrt{E^2 -  {L^2 e^\nu
\over r^2} } \qquad { d t \over d \lambda} = E e^{-\nu},
\end{equation}
which can in principle be solved for $t(\lambda),~r(\lambda)$ and
inverted to obtain $\lambda =\lambda ( r ) $. For the interesting case of 
time-independent metric this gives
\begin{equation}
\Phi\up m  = - {m^2\over 2 E}
\int dr { e^{(\nu +\gamma)/2 } \over  \sqrt{1 -  {L^2 e^\nu
\over E^2 r^2} } } ,
\label{spher.sym}
\end{equation}
It appears possible for this to retain a gravitational contribution
even when $ L=0 $ (radial geodesics). To determine whether this is in
fact the case I will consider
the case of an ideal fluid of energy density $\rho $ and pressure $p$
both of which vanish for $ r > R $.
The solution to Einstein's equations gives~\cite{weinb}
\begin{eqnarray}
\gamma &=& \ln [ 1 - 2 m(r) G/r], \qquad m(r) = \int_0^r 4 \pi u^2
\rho(u) du \cr
\nu &=& - 2 G \int_r^\infty  du { m(u) + 4 \pi u^3 p(u) \over u[ u - 2
m(u) G ] } ,
\end{eqnarray}
 where $G$ denotes Newton's constant;
for $ r > R $ $\nu $ and $ \gamma $ reduce to their Schwarzschild expressions.
To first order in $G$ I find
\begin{equation}
e^{(\nu+\gamma)/2} = 1 + 4 \pi G \int_r^\infty du \; u [ \rho(u) + p(u)]
\end{equation} 
which, when substituted into (\ref{spher.sym}) with $L=0$ gives the
phase for radial motion.
If the initial point of the geodesic $r_i $ satisfies $ r_i < R $ and
final point lies beyond $R$ then a simple
estimate gives $ \Phi \up m \sim (m^2/2 E)(r + c r_g ) $ where $r_g$
denotes the gravitational radius of the matter distribution, and $c$ is a
numerical constant $ \lsim O(1) $ that depends on the detailed form of 
$ \rho $ and $p$. Note that, just as in (\ref{kerr.a},\ref{kerr.b}), 
the leading gravitational 
contribution to the phase depends on the initial point and the details
of the metric, but not on $r$; if expressed in terms of $ \ell $, however,
the phase again acquires an $ m^2 \ln \ell$ contribution.

\paragraph{Weak gravitational interactions}
For the case of a general weak metric it is possible to obtain a rather
simple expression for the phase $ \Phi\up m $. Expanding $S_0$
(cf. (\ref{action})) in powers of the metric perturbation,
\begin{equation}
g_{\mu\nu} = \eta_{\mu\nu}+ h_{\mu\nu}; \qquad
S_0 = S_0\up0 + S_0\up1 + \cdots;  \quad S_0\up1 = O(h),
\end{equation}
(where $ \eta$ denotes the flat-space metric) and defining
\begin{equation}
p_\mu = - \partial_\mu S_0\up0 .
\end{equation}
I find $ p^2 =0 $ and
\begin{eqnarray}
p^\mu \partial_\mu S_0\up0 = - {1\over2} h_{\mu\nu} p^\mu p^\nu,
\end{eqnarray}
where indices are raised and lowered using the flat metric $ \eta $.

It proves convenient  to choose coordinates such that
\begin{equation}
p^\mu=E (1,1,0,0),
\end{equation}
then the equation for $ S_0\up1 $ becomes
\begin{equation}
{\partial S_0\up1 \over \partial x_+ } = - {1\over2} ( h_{00}+2h_{01}+h_{11}
), \qquad x_+ = {x^0+x^1 \over2},
\end{equation}
so that 
\begin{equation}
S_0 = - p \cdot x - {1\over2}E \int d x_+ ( h_{00}+2h_{01}+h_{11} ) +
\cdots .
\end{equation}
From this and the Hamilton-Jacobi equation I find
\begin{equation}
{d x^\mu \over d \lambda } = p^\mu + 
{1\over2} E \partial^\mu \int d x_+ ( h_{00}+2h_{01}+h_{11} ) -
h^{\mu\nu} p_\nu + \cdots,
\end{equation} and, in particular,
\begin{equation}
{d x^1\over d \lambda}  = E - {1\over2} E \partial_1 \int d x_+ ( h_{00}+2h_{01}+h_{11} ) + 
E(h_{11}+h_{10})+ \cdots.
\end{equation}

For a time-independent metric the above expression simplifies to
\begin{equation}
{d x^1\over d \lambda}  = E - {1\over2}E ( h_{00}-h_{11} ) ,
\end{equation}
whence $ \lambda E = x + {1\over2} \int dx ( h_{00}-h_{11} ) $ and
\begin{equation}
\Phi\up m = - { m^2\over2 E} \left[x + {1\over2} \int dx ( h_{00}-h_{11} ) + \cdots
\right],
\end{equation}
where the phase difference must be obtained by evaluating the quantity
in brackets at two different points on a geodesic with the line integral
being along the same geodesic.

The gravitational terms vanish if the geodesic has zero angular momentum
for the case of a distant (localized) matter distribution since in this
case~\cite{ll} $
h_{00} = - r_g/r $ and $ h_{ij} = - r_g x^i x^j/r^3 $ (in Cartesian
coordinates). Moreover, since $\Phi \up m $ is independent of $ h_{0i}
$, the dependence on the angular momenta will be quadratic.
This result is
not accidental: linear terms are absent since the phase is a scalar 
and the angular momentum is a pseudovector.
In terms of the invariant length 
\begin{equation}
\Phi\up m = -{m^2\over2E}\left[\ell + {1\over2} \int d\ell\; h_{00}
+ \cdots\right]
\end{equation}
which again has a linear+logarithmic dependence on $ \ell $.

\paragraph{Discussion}
\label{sect:discussion}
The leading mass dependence of the phase of a general wave function is,
within the WKB approximation, proportional to the affine parameter along
the geodesic followed by the wave packets and may
contain explicit gravitationally-induced contributions. 
The expression for the 
gravitationally-induced phase depends on the initial conditions
assumed for the system and on the choice of physical variables.
 This last point is particularly delicate since the
separation of $ \Phi\up  m $ into gravitational and flat-space contributions
is sensitive to the quantity used as physical distance. The
results presented in ~\cite{al,mot,oth} are
consistent with each other provided these points are taken into account.

In all interesting cases, however, 
the uncertainty in the distances to regions of strong gravitational field are
too large for these effects to be unambiguously isolated experimentally. In terms of
the radial coordinate $r$ (defined at large distances using the
Schwarzschild metric) and for the case of non-zero angular momentum,
$ \Phi\up m \sim -
( m^2/2E )(r + r_0^2/r_i)$ where $r_i$ denotes the initial point of the
geodesic and $r_0$ is a characteristic length of the problem (such as the
geodesic impact parameter or the gravitational angular momentum per
unit mass) such that the angular momenta are of order $ r_0 E $. 
In all cases considered $ r_0 < r_g < r_i $ so that  $ \Phi\up m \sim -(m^2/2E)(r + r_g) $. 
These gravitational effects (for the case where
the flat-space phase is defined as $ -m^2 r /2E $)
can be isolated experimentally only
when $r$ is known to an accuracy better than $r_g$, and this is not usually
attainable.

In terms of the invariant length $ \ell $ the phase takes 
the form $ \Phi\up m \simeq (m^2/2E)[\ell + (r_g/2)
\ln(\ell/r_g) ] $ with a more significant dependence on $ r_g$. 
Nonetheless the precision in $ \ell $ required to extract the
gravitational contribution (where
the flat-space phase is now defined as $ m^2 \ell  /2E $) is still not
available. For example for a $10 M_\odot$ supernova 
$ 10^5 $ light years away, $ \ell $ should be known better
to a precision better than $2,000$ km or $10^{-13}~\%$; for a $10^9 M_\odot$
supermassive quasar $10^8$ light years away, $ \ell $ should be
known to a precision better than $10^{11}$ km or $10^{-10}~\%$.
Despite the logarithmic enhancement the current experimental sensitivity
precludes an accurate discrimination of the gravitational contributions
for realistic situations.

\bigskip\bigskip

\acknowledgements
The author would like to thank S. Pakvasa and D. Ahluwalia for
illuminating comments. This work was supported in part through funds
provided by the U.S. Department of Energy under Grant No. DE-FG03-94ER40837.

\end{document}